\documentclass{article}

\font\mybold=cmmib10
\chardef\Myiota="13

\font\mybold=cmmib10
\chardef\Myxi="18
\newcommand{\boldxi}{\hbox{\mybold\Myxi}}
\font\mybold=cmmib10
\chardef\Myiota="13
\newcommand{\X}{\bf X}
 \newcommand{\thorn}{\hbox{\rm I}\kern-0.32em\raise0.35ex\hbox{\it o}}
 \newcommand{\edth}{\hbox{$\partial$\kern-0.25em\raise0.6ex\hbox{\rm\char'40}}}

 \newcommand{\bedth}{{\pmb{\edth}}}
 \newcommand{\bthorn}{{\pmb \thorn}}
 \newcommand{\bthornp}{{\bthorn^\prime}}
 \newcommand{\bedthp}{{\bedth^\prime}}

 \newcommand{\Ph}{\nthorn}
 
 \def\pmb#1{\setbox0=\hbox{#1}  \kern-.025em\copy0\kern-\wd0
 \kern.05em\copy0\kern-\wd0
  \kern-0.025em\raise.0433em\box0 }

 \newcommand{\nthorn}{{\pmb\thorn}}

\begin{document}

\title{Symmetry analysis of radiative spacetimes with a null isotropy using GHP formalism.}

\author{ S. Brian Edgar\footnote{Department of Mathematics, Link\"{o}pings universitet, Link\"{o}ping, Sweden},
Michael Bradley\footnote{Department of Physics, Ume{\aa} universitet, Ume{\aa}, Sweden} \footnote{email: michael.bradley@physics.umu.se}
 and  M. Piedade Machado Ramos\footnote{Departamento de Matem\' atica e Aplica\c{c}\~oes, Guimar${\hbox {a}}$es, Universidade do Minho, Portugal} \footnote{email: mpr@mct.uminho.pt}
}

\maketitle

\begin{abstract}
A complete and simple invariant classification of the conformally flat pure radiation metrics with a negative cosmological constant that were obtained by integration using the generalised invariant formalism is presented. We show equivalence between these metrics and the corresponding type O subclass of the more general spacetime studied by Siklos. The classification procedure indicates that the metrics possess a one degree of null isotropy freedom which has very interesting repercussions in the symmetry analysis. The Killing and homothetic vector analysis in GHP formalism is then generalised to this case were there is only one null direction defined geometrically. We determine the existing Killing vectors for the different subclasses that arise in the classification and compare these results to those obtained in the symmetry analysis performed by Siklos for a larger class of metrics with Ricci tensor representing a pure radiation field and a negative cosmological constant.  It is also shown that there are no homothetic Killing vectors present.

\end{abstract}

\section{Introduction}

The success and simplicity of the study of a particular spacetime is influenced by the coordinate system used,
and of course by the type of analysis required.  Analyses of the familiar spacetimes have usually been
motivated by physical considerations, and the coordinate systems employed reflect this. On the other hand, if a more
mathematical analysis of a spacetime is required the original coordinates may not be the most convenient.
Recently, \cite{edram4}, we have investigated in detail the invariant classification and symmetry analysis of the
conformally flat pure radiation spacetimes with zero cosmological constant \cite{edludL}, \cite{edlud2}, \cite{edvic}; we demonstrated
how  these procedures were much simpler and transparent using the version of the metric generated by the Generalised
Invariant Formalism
(GIF) integration procedure
in \cite{edvic},  as compared to the version given in more familiar Kundt-type coordinates in \cite{skea1}. A symmetry analysis of these metrics can also be found in \cite{barnes1}.

The same integration procedure as used in \cite{edvic} --- of generating metrics via GIF in tetrads and coordinates which are {\it intrinsic} (as far as possible)  ---  has also been used
to find  the conformally flat pure radiation metrics with  non-zero cosmological constant, in two other
papers \cite{edram2}, \cite{edram3}.

These three classes of spacetimes in  \cite{edvic}, \cite{edram2}, \cite{edram3}
together comprise  the complete family of conformally flat  pure radiation Kundt spacetimes with a cosmological constant,
and Podolsk\'y and Prikryl \cite{pod2}
have recently shown explicitly how the versions of the metrics given in \cite{edvic}, \cite{edram2}, \cite{edram3}  relate to the metric of the more general
family of conformally flat (and type N) Kundt spacetimes which are either vacuum or pure radiation, with a
cosmological constant, and were originally given by
Ozsv\'ath, Robinson and R\'ozga in \cite{ozs}.

There are a number of interesting aspects to this particular class of  spacetimes in \cite{edram2}, and we expect to get some more insight into their properties, and also into how the invariant classification and symmetry analysis in GIF handles such spacetimes. In particular, this class of spacetimes has one degree of null rotation isotropy freedom. An efficient way of investigating Killing and homothetic vectors of metrics obtained by integration in the GHP formalism is described in \cite{edlud3} and \cite{edlud10}. Of course this method assumes that two null directions are singled out and that these constitute the intrinsic GHP tetrad. We show how one can generalise this analysis to the case where only one null direction is defined geometrically. In doing so we are able to obtain the existing Killing vectors.

\subsection{Equivalence problem and invariant classification}

The equivalence problem is the problem of determining whether the
metrics of two spacetimes are locally equivalent, and the original
contribution of Cartan \cite{car} directed attention to the
Riemann tensor and its covariant derivatives up to $(q+1)$th order,
${\cal R}^{q+1}, $ calculated in a particular frame.

 In going from ${\cal R}^q$ to ${\cal
R}^{q+1}$ for a particular spacetime, if  there is no new functionally independent Cartan scalar
invariant  and  ${\cal R}^q$ and ${\cal R}^{q+1}$ have equal isotropy group, then all the local
information that can be obtained  about the spacetime is contained
in the set  $ {\cal R}^{q+1}$. The set $ {\cal R}^{q+1}$ is
called the {\it Cartan scalar invariants}  and provide the information
for an {\it invariant classification} of the spacetime.

A  practical method for invariant classification was developed by
Karlhede \cite{karl}, using fixed frames. In this algorithm the
number of functionally independent quantities is kept as small as
possible at each step by putting successively the curvature and its
covariant derivatives into canonical form, and only permitting those
frame changes which preserve the canonical form. In practice, rather
than work directly with the Riemann tensor and its covariant
derivatives, it is convenient to use decompositions of their spinor
equivalents; a minimal set of such spinors has been obtained in
\cite{maccullum}. The frame components of ${\cal R}^{q+1}$  in the
canonical
 frame  fixed by the Karlhede algorithm will be called the {\it Cartan-Karlhede  scalar
invariants}\footnote{In the literature these are simply
called the {\it Cartan scalar invariants}, but we wish to
distinguish these invariants found using the Karlhede algorithm from
other invariants.}.

The GHP formalism is particularly efficient when a pair of  intrinsic spinors has been identified by the geometry, and calculations  can be carried out in a spin- and boost-weighted scalar  formalism, rather than restricting the spin and boost freedom in an ad hoc manner.
Within this formalism an invariant classification procedure was discussed in \cite{cdv1}, \cite{cdv2} and \cite{cdI}, and applied to
Petrov Type D spacetimes.
 The (weighted) {\it GHP Cartan scalar  invariants}
 used for the invariant classification are closely related to, but not identical with the
 Cartan-Karlhede scalar invariants used in the Karlhede algorithm: see \cite{edram4} for a full discussion.

The GIF is particularly efficient  when one intrinsic spinor has been identified by the geometry, and calculations for an invariant classification can be carried out  in a  spin- and boost-weighted and null rotation invariant spinor formalism, rather than restricting the spin, boost and null rotation freedom of the frame in an ad hoc manner. Within this formalism an invariant classification procedure has been  discussed, and applied to Petrov Type N spaces in \cite{maria4}, \cite{maria3}, and to conformally flat pure radiation spaces in \cite{edvic}, \cite{edram4}.
The  information
for this invariant classification is carried by the {\it Cartan spinor invariants}.  However, as soon as a
second invariant spinor has been supplied within the GIF calculations, transfer can be
made to the  GHP formalism, and work continued
in this simpler scalar formalism using  GHP Cartan scalar invariants, which are scalar versions of the
Cartan spinor invariants.  In \cite{edvic}, \cite{edram4},  the spacetimes under investigation supplied a second intrinsic spinor  at second order of the spinor invariants, and this permitted the transfer to the GHP formalism in which the invariant classification was completed.

On the other hand, there are situations (corresponding to   null rotation isotropy  associated with the second null direction) where the GIF does not supply a  second intrinsic spinor. In such cases we need to complete the invariant classification procedure
in GIF; this is the case of the metrics under discussion in this paper.

\subsection{Killing vector analysis in GHP formalism}

Within tetrad formalisms there is a standard method for finding Killing vectors for a given spacetime: integrate  the tetrad version of the Killing equations.
This procedure is usually long and complicated, as well as inefficient, because there is usually considerable redundancy within the Killing equations due to the fact that there is likely to be a relationship  between the Killing vector(s) and the tetrad; but this simplifying relationship  is not usually exploited in the explicit calculations.

It is shown in \cite{edlud3} that  when a spacetime has been calculated within the GHP formalism, then the symmetry investigations can be considerably shortened. The integration procedure, and the symmetry analysis within GHP formalism is explained in detail, and illustrated with applications to a variety of metrics in \cite{edlud3}, \cite{edlud10}. Here we summarise the crucial definitions\footnote{Some of these   are more precise   definitions of  ideas and terms used by
Held    in \cite{held3},\cite{held4}.
}, theorems and procedure,  quoting \cite{edlud3}.

\noindent
{\bf Definitions.}

\noindent
$\bullet$ A \underbar{\it (standard) GHP tetrad,} $Z_m{}^\mu, \ m=1,2,3,4,$ is a class  of the
usual null tetrads
$ (\bf {Z_1 =l, Z_2 =n, Z_3 =m, Z_4 =\bar m}) $ with the directions of the two real null
vectors $\bf {l, n, } $ chosen, but possessing the  two-dimensional gauge
freedom of spin and boost transformations.
\smallskip

\noindent
$\bullet$ A  scalar quantity
$\eta$ of the GHP formalism   is said to have {\it GHP weight}
$\{p,q\}$ if under this tetrad gauge transformation, it transforms as
\begin{equation}
\eta
\rightarrow
\lambda^{(p+q)/2}e^{i(p-q)\theta/2} \eta.
\end{equation}
We will refer to such a scalar as being {\it (non-trivially) weighted} if $p\ne \pm q$.

\smallskip

\noindent
$\bullet$ The \underbar{\it
GHP scalars
} (with respect to a standard GHP tetrad)  are defined
   to be all the well-behaved GHP
spin coefficients
$\rho,\sigma,  \ldots$,  all the Riemann tensor components, $\Psi_i, \Phi_{ij},
\Lambda$,  all the  GHP derivatives of the spin coefficients
and of the Riemann tensor components   $\thorn
 \rho, \edth \thorn' \Psi_4, \ldots$, together with
properly weighted functional combinations of all of these.
 \smallskip

\noindent
$\bullet$ The
\underbar{\it Lie scalars} (of the GHP formalism for the  vector $\boldxi $)
are the four  scalars   $\xi _m(=\xi ^\mu
Z_{m\mu}), \  \ m=1,2,3,4, $ \  which are the GHP tetrad
components of the  vector $\boldxi $, as well as all  their GHP derivatives $
 \thorn \xi_m , \thorn' \delta \xi_m . \ . \ . \ .$, together with properly weighted
functional combinations of all of these.
 \smallskip

\noindent
$\bullet$  \underbar{\it Complementary scalars}  are any other scalars --- other than GHP scalars and Lie scalars --- which are introduced into the calculations.

Such scalars  will be always introduced in a manner such that they have good
weight.

\smallskip

\noindent
$\bullet$ The \underbar{\it GHP Lie derivative operator} (with respect to a vector
$\boldxi $)
$\hbox{\L}_\xi $
is defined in a spacetime with  GHP tetrad $ (\bf {l, n,
m,
\bar m}) $ by
\begin{equation}\label{GHPL}
\hbox{\L}_\xi  =\pounds_\xi  -({p\over 2}+{q\over 2})n_\mu \pounds_\xi  l^\mu +
({p\over 2}-{ q\over 2})
\bar{m}_\mu \pounds_\xi  m^\mu
\end{equation}
where  ${\pounds}_\xi $ is the usual Lie derivative with
respect to a vector field  ${\bf \boldxi }$.

\noindent
The operator $\hbox{\L}_\xi $ is
 well behaved under spin and boost transformations
and has weight $\{0,0\}$.; furthermore,  $\hbox{\L}_{\xi }$ obviously is equivalent  to  the usual Lie operator $\pounds_\xi $ when acting on a $\{0,0\}$ quantity.

\smallskip
\noindent
$\bullet$
A weighted scalar or tensor quantity $T^{\mu\nu
...}{}_{\alpha\beta ...}$  is \underbar{\it GHP Lie derived} if
\begin{equation}
\hbox{\L}_\xi  (T^{\mu\nu ...}{}_{\alpha\beta ...}) =0.
\end{equation}

\medskip

\smallskip

\noindent
$\bullet$ An \underbar{\it
intrinsic  GHP  tetrad}
is a standard GHP tetrad  that has been chosen so that  the direction
of each of the two real null vectors
${\bf l}, {\bf n}$
    is
determined by an invariantly defined tensor field of the  Riemann
tensor and its covariant derivatives (to whichever order is necessary), e.g. by  the principal
null directions of the Weyl tensor, or the velocity vector of a perfect fluid.

\smallskip
\noindent
$\bullet$ The \underbar{\it intrinsic GHP scalars} are the GHP scalars
 defined with respect to an intrinsic GHP tetrad.

\smallskip
\noindent
$\bullet$
A set
 of $r (\le 4)$ $\{0,0\}$  real intrinsic GHP scalars which are functionally
independent will be called
\underbar{\it intrinsic GHP coordinates}; $(4-r)$ $\{0,0\}$
real complementary scalars which are functionally independent of this set, and of each other,
can be used as  \underbar{\it  complementary  coordinates.}

\smallskip

\noindent
{\bf Results}

\noindent
The basic results (Lemmas 2 and 3 in \cite{edlud3}) giving the properties of the GHP Lie derivative operator $\hbox{\L}_\xi$   are as follows.

\noindent
\underbar{Lemma 2.}(\cite{edlud3})(a) In any spacetime the vector field $\boldxi$ is a Killing vector field, if and only if, there exists a GHP tetrad ${Z_m}^\mu$ which is GHP Lie derived by the associated GHP Lie derivative operator $\hbox{\L}_\xi$,

\begin{equation}
\hbox{\L}_\xi{Z_m}^\mu=0
\label{L2a}
\end{equation}

\medskip
\noindent
(b) Furthermore, all the GHP scalars constructed from this particular GHP tetrad are GHP Lie derived with respect to this Killing vector field $\boldxi$,

\begin{eqnarray}
&&\hbox{\L}_\xi\rho = \hbox{\L}_\xi\sigma=...=0,\nonumber \\
&&\hbox{\L}_\xi\Psi_{i}= \hbox{\L}_\xi\Phi_{ij}=\hbox{\L}_\xi\Lambda =0,\nonumber\\
&&\hbox{\L}_\xi\thorn \rho = ....=0, \hbox{\L}_\xi\edth\thorn^\prime \Psi_0 =...=0,\nonumber\\
&&.......    .
\label{liederivedscalars}
\end{eqnarray}

\medskip
\noindent
The Lie scalars of this GHP tetrad are also GHP Lie derived with respect to this Killing vector field,

\begin{eqnarray}
&&\hbox{\L}_\xi\boldxi_m =0=\hbox{\L}_\xi\boldxi^m,\nonumber \\
&&\hbox{\L}_\xi\thorn \boldxi_m = \hbox{\L}_\xi\thorn^\prime\edth \boldxi_m =...=0,\nonumber\\
&&.......    .
\label{liederivedscalars*}
\end{eqnarray}

\medskip
\noindent
\underbar{Lemma 3.}(\cite{edlud3})(a) In any spacetime, the vector field $\boldxi$ is a Killing vector field, if and only if, there exists a GHP tetrad ${Z_m}^\mu$ with associated GHP Lie derivative operator ${\hbox{\L}_\xi}$ which commutes with the GHP operators $(\thorn , \thorn^\prime , \edth ,\edth^\prime )$ when acting on an arbitrary
$\{0,0\}$ weighted scalar $\eta$,

\begin{equation}
 \Bigl[ \hbox{\L}_\xi,
\thorn\Bigr]\eta=0, \label{comm}
\end{equation}
 and its companion equations.

\noindent
 (b) Furthermore, with respect to any Killing vector $\boldxi$, the commutator equations (\ref{comm}) are also satisfied when $\eta$ is any $\{p,q\}$ weighted scalar, or indeed any tensor.

\noindent
The following theorems (Theorems 4 and 5 in \cite{edlud3})  provide the key tools, when the spacetime under consideration contains an intrinsic GHP tetrad.

\noindent
\underbar{Theorem 4.}(\cite{edlud3})  (a)\  In a spacetime containing    an  intrinsic GHP
tetrad, the vector field  $\boldxi$  is a Killing vector field  if,  and only if,   the intrinsic
GHP tetrad is GHP  Lie derived by its associated GHP Lie derivative operator $ {\hbox{\L}_\xi}$.

\noindent
(b) Furthermore, all the intrinsic GHP scalars and Lie scalars constructed from the intrinsic GHP tetrad are also Lie derived with respect to this Killing vector field.

\smallskip
\noindent
\underbar{Theorem 5.}(\cite{edlud3})\ (a) \ In a spacetime containing   an
intrinsic GHP tetrad,   the vector field  $\boldxi$ is a Killing vector field  if, and only if,
 the associated GHP Lie derivative $ {\hbox{\L}_\xi}$
commutes with the GHP operators
$(\thorn, \thorn', \edth, \edth')$ when acting on an arbitrary $\{0,0\}$ weighted scalar $\eta$ (equivalently four functionally independent $\{0,0\}$ weighted scalars), that is (\ref{comm}) and its three companion equations are satisfied when acting on such a scalar.

\noindent
 (b)\ Furthermore, in the presence of the Killing vector $\boldxi$ the Lie-GHP commutator equations (\ref{comm}) are
also satisfied when $\eta$ is any $\{p,q\}$ weighted scalar, or indeed any tensor.

\smallskip
\noindent
{\bf Principles and Procedures }

\noindent
The fundamental principles of the GHP integration
procedure are discussed in detail in  \cite{ed2}, \cite{edlud2}, \cite{edlud1}, and can be summarised as :

\noindent
$\bullet $ to  work as much as possible with respect to an intrinsic GHP tetrad, and use intrinsic GHP scalars
constructed from that tetrad.

\noindent
$\bullet$ to use only $\{0,0\}$ scalars  as coordinates.

\noindent
Then the two theorems above tell us that if we carry out a GHP integration in an intrinsic GHP tetrad  then we can
more readily draw conclusions from our results regarding the existence of Killing
vector fields. By applying the commutators (\ref{comm}) and its three companions enables the Killing vectors $\boldxi$ to be calculated.

\noindent
In particular, in a symmetry analysis of a metric obtained in this way, simplification occurs because many of the scalars will be GHP Lie derived; also, when the Lie-GHP commutators (which replace the Killing equations) are applied to the four coordinates (functionally independent zero weighted  scalars) obtained during the GHP integration procedure, often significant parts of the calculations are trivial.

\subsection{Killing vector analysis in GHP formalism: with null rotation isotropy}

The procedure to determine Killing vectors --- devised in \cite{edlud3}, and summarised above ---    assumed that the spacetime  under discussion was one where two null directions were  defined geometrically, and these constituted  the intrinsic GHP tetrad. In certain cases a second null direction is not defined geometrically (or at least such a direction is not easy to find); however,
it is easy to generalise the procedure to the case where  there appears to be only one null direction defined geometrically.

\medskip
In such a case we exploit the earlier results --- Lemma 2 and 3 --- quoted above.

\medskip
From Lemma 2 we know that, for any Killing vector present, there must be a second null direction which is GHP Lie derived  \cite{edlud3} (but not necessarily intrinsic), and so we change our calculations into  that (still unidentified) tetrad given by

\begin{equation}\widehat {\bf l} = {\bf l}; \qquad \widehat {\bf m} = {\bf m} + z {\bf l}; \qquad \widehat {\bf n} = {\bf n} + \bar {z} {\bf m}+{z}  \bar{\bf m} + z\bar z {\bf l}
\label{newbasis}
\end{equation}
where we have assumed that, in the original tetrad, the first null direction ${\bf l}$ is defined geometrically, and is kept unchanged. A particular value of the complex null rotation parameter $z$, which has weight  $\{0,-2\}$ will correspond to the tetrad which is GHP Lie derived.
Hence, with respect to that tetrad $\widehat {\bf l}, \widehat {\bf m}, \widehat{\bar{\bf m}}, \widehat {\bf n}$,
we use the GHP Lie operator \begin{equation}\label{GHPLhat}
\widehat{\hbox{\L}}_\xi  =\pounds_\xi  -({p\over 2}+{q\over 2})\widehat n_\mu {\cal L}_\xi  \widehat l^\mu +
({p\over 2}-{ q\over 2})
\widehat {\bar{m}}_\mu {\cal L}_\xi \widehat m^\mu
\end{equation}
where  ${\cal L}_\xi $ is the usual Lie derivative with
respect to a vector field  ${\bf \boldxi }$;
and in the GHP-Lie commutators  the associated operators $\widehat{\thorn}, \widehat{\thorn'}, \widehat{\edth}, \widehat{\edth'}, $ will be used,

\begin{equation}
 \Bigl[\widehat{\hbox{\L}}_\xi,
\widehat {\thorn}\Bigr]\eta=0, \label{commhat}
\end{equation}
with the companion equations.

 In such a situation the commutators are solved for the Killing vectors $\boldxi$, together with the null rotation parameter $z$  (which will not necessarily be the same for each Killing vector)\footnote{See \cite{edlud11} for a related discussion of isotropy freedom in the NP formalism.}.

In the spacetime under consideration in this paper, it becomes apparent in Section 3, that the second null direction is only defined geometrically up to one (imaginary) degree of isotropy, $z=i\varepsilon$, where $\varepsilon$ is real, and so the modification just described (involving a new tetrad) is necessary for a symmetry analysis.

\subsection{Outline}

In the next section we give the various versions of the spacetime which is being considered. In Section \ref{icint} we  carry out a detailed invariant classification and symmetry analysis on the version of the metric which was obtained by a GIF/GHP derivation resulting in intrinsic tetrad and coordinates (as far as possible).

In Appendix \ref{classint} we compare these results with the usual Karlhede classification using CLASSI for the same coordinate version of this class of spacetimes; in Appendix \ref{classsik} we compare with the classification  in the Siklos coordinate version .

\section{The metric}

In \cite{edram2}, following integration in the GIF/GHP formalism the null tetrad is given in coordinates
 $ r,  n,m,  b$ by
\begin{eqnarray}\nonumber
{\bf{l}}&=&\frac{1}{\cal{Q}}\left(0,m^{3/2},0,0\right), \quad {\bf{n}} = {\cal{Q}} \left(m^{1/2}, \frac{V m^{1/2}}{2}, \nu_4 m^{3/2}, \nu_4 b m^{1/2}\right), \\\label{frame}
{\bf{m}}&=&{\cal{P}}\left(0,0,\lambda m, i\lambda m \right) \, ,
\end{eqnarray}
with $V$  given by
\begin{equation}\label{Usim3}
V  ={3}  n\nu_4 -  {\nu_5}( b^2+
 m^2)-{\nu_6 b}  -\frac{ m}{\lambda^2} +\nu_3
\end{equation}
where $\nu_3\equiv\nu_3( r), \nu_4\equiv\nu_4( r), \nu_5\equiv\nu_5( r),
\nu_6\equiv\nu_6( r), $ all are completely arbitrary functions
of $ r$. The cosmological
constant $\Lambda =-6\lambda^2$, and clearly we could replace $6\lambda^2$ with $-\Lambda$ in the
expression for $V$, but we choose to retain the former to emphasise that for this spacetime the
cosmological constant must be negative (a positive $\Lambda$ would give a different signature).
${\cal P}$ and ${\cal Q}$ are weighted scalars which represent the spin and boost freedom; ${\cal Q}$ has weight $\{-1,-1\}$ and is real,  while  complex ${\cal P}$ has weight $\{1,-1\}$ and satisfies ${\cal P}\bar {\cal P}=1$.
 (Note the slightly different definitions for  ${\cal P}, {\cal Q}$ in \cite{edram2}, compared to $P,Q$ in \cite{edram3}, \cite{edvic} \cite{edram4}.)
The metric $g^{\mu\nu}=\eta^{ab}Z_a^\mu Z_b^\nu$  is hence given by
\begin{equation}
g^{ij}=  { m^2}\left(
\begin{array}{lccr}
0 &1& 0 & 0
\cr 1&
V &  m\nu_4&
 b \nu_4
\cr 0 &   m\nu_4&-{2}\lambda^2 & 0
 \cr 0 &
 b  \nu_4
 & 0 &-{2}\lambda^2
\end{array}
\right)  \label{metric1up}
\end{equation}
or equivalently,
\begin{eqnarray}\label{metric1}
&&\hbox{d}s^2= \Bigl(-(2\lambda^2V+\nu_4^2(m^2+b^2))\hbox{d}r^2 + 4\lambda^2\hbox{d}r\hbox{d}n\nonumber\\ &&
+2m\nu_4\hbox{d}r\hbox{d}m +2b\nu_4\hbox{d}r\hbox{d}b-\hbox{d}m^2-\hbox{d}b^2\Bigr)/2\lambda^2m^2 \, .
\label{metric1down}
\end{eqnarray}
Podolsk\'y and Prikryl \cite{pod2} have pointed out that these spaces are a subclass of the spacetimes
originally given
 by  Siklos \cite{sik}
\begin{equation}
\hbox{d}s^2=  -\bigl( \hbox{d}x^2+\hbox{d}y^2+2\hbox{d}u\hbox{d}v+H(x,y,u)\hbox{d}u^2\bigr)\big/ 2\lambda^2 x^2\label{metric1.1down}
\end{equation}
with
\begin{equation}\label{Hu}
H(x,y,u) = A (x^2+y^2)+Bx +Cy+D
\end{equation}
where $A\equiv A(u), B\equiv B(u), C\equiv C(u),D\equiv D(u)$ are arbitrary functions of $u$.
In fact, with the coordinate transformation
\begin{eqnarray}\nonumber\label{coordtransform}
u&=&2\lambda^2\int^r\exp{\left(\frac{3}{2}\int^s\nu_4(t) dt\right)}ds \, , \quad x=m \, , \quad y=b\\
v&=&-\exp{\left(-\frac{3}{2}\int^r\nu_4(s) ds\right)}\left(n+\frac{\nu_4}{4\lambda^2}(m^2+b^2)\right) \, ,
\end{eqnarray}
which will be more useful for our purposes than the one used in \cite{pod2}, the metric form (\ref{metric1.1down}) is obtained
with the arbitrary functions of $u$ given by
\footnote{The function $B$ can be made positive by instead using a coordinate transformation with $x=-m$,
but note that $B$ cannot be made equal to zero, which would correspond to the anti de-Sitter spacetime.}
\begin{eqnarray}\nonumber
A(u)&=&-\frac{\exp{\left(-3\int^r\nu_4(s)ds\right)}}{4\lambda^4}\left(2\lambda^2\nu_5-
\nu_4^\prime+\frac{1}{2}\nu_4^2\right)\, ,\\\nonumber
\quad B(u)&=&-\frac{\exp{\left(-3\int^r\nu_4(s)ds\right)}}{2\lambda^4}\, , \quad
C(u)=-\frac{\exp{\left(-3\int^r\nu_4(s)ds\right)}}{2\lambda^2}\nu_6\, ,\\\label{ABCD}
D(u)&=&\frac{\exp{\left(-3\int^r\nu_4(s)ds\right)}}{2\lambda^2}\nu_3\, ,
\end{eqnarray}
where prime denotes differentiation with respect to $r$ and the relation between $r$ and $u$ is given by (\ref{coordtransform}). The form (\ref{Hu}) of $H(x,y,u)$ also
follows from the requirement that the Weyl tensor should vanish.

Siklos's larger class, which  includes  type N  spaces,
has  been analysed in \cite{pod1}.

As noted in the previous section, the metric (\ref{metric1down}) describes a spacetime with {\it non-zero} pure radiation, $\Phi_{22}\ne 0$, and {\it non-zero} cosmological constant, $\Lambda\ne 0$; on the other hand, the metric (\ref{metric1.1down}) includes the limiting cases of {\it zero} pure radiation (Einstein space), but does not include the case of   zero cosmological constant.

Podolsk\'y and Prikryl \cite{pod2} have also given the explicit Kundt form

\begin{equation}
\hbox{d}s^2=  -\frac{2}{P^2}\hbox{d}\zeta\hbox{d}\bar{\zeta}+\frac{2Q^2}{P^2}\hbox{d}u\hbox{d}v-\bigl( \kappa\frac{Q^2}{P^2}v^2-\frac{(Q^2)_{,u}}{P^2}v-\frac{Q}{P}H\bigr)\hbox{d}u^2\label{metricKundt}
\end{equation}
where
\begin{equation}
P=1+\Lambda\zeta\bar{\zeta}/6\,\,\,\,\,\,\,\,\,\,\,\,\,\,\, Q=(1-6\Lambda\zeta\bar{\zeta})\alpha +\bar{\beta}\zeta +\beta\bar{\zeta}\nonumber
\end{equation}
and
\begin{equation}
\kappa =2\beta\bar{\beta}+\lambda\alpha^2/3\nonumber
\end{equation}
$\alpha (u)$, $\beta (u)$ are functions of $u$, and
\begin{equation}
H(\zeta ,\bar{\zeta}, u)=\frac{\mathcal{A}(u)+\bar{\mathcal{B}}(u)\zeta + \mathcal{B}(u)\bar{\zeta}+\mathcal{C}(u)\zeta\bar{\zeta}}{1+\lambda\zeta\bar{\zeta}/6}
\end{equation}
where $\mathcal{A}(u)$, $\mathcal{B}(u)$ and $\mathcal{C}(u)$ are arbitrary functions of $u$ with $\mathcal{A}(u)$ and  $\mathcal{C}(u)$ real. This version includes both the limiting cases of zero pure radiation (Einstein space), and zero cosmological constant. The class was orignially given by Ozsv\'ath, Robinson and R\'ozga in \cite{ozs}.

\section {The invariant classification using 
 Cartan invariants in GIF/GHP generated coordinates}\label{icint}

\subsection {The spinor Cartan invariants}\label{GIFclass}

Some explicit expressions for the
Cartan spinor invariants in the Karlhede classification were quoted in \cite{edram2}, and these are
now repeated and completed.

\noindent
\underbar{At zeroth order,} there are only the two Cartan spinor invariant
\begin{equation}\label{zeroth}
 \Phi  = {\cal Q}^2 \label{zero}, \qquad \qquad \Lambda = -\lambda^2
\end{equation}
\underbar{At first order,} there are the eight Cartan spinor invariants
\begin{eqnarray}  \label{first}
\bthorn {\Phi} &=& 0, \quad
\bedth {\Phi}=  \lambda {{\cal Q}}^2{{\cal P}}, \quad
\bedthp {\Phi} = \lambda {{\cal Q}}^2\overline{{\cal P}}, \quad
\bthornp {\Phi} =- 3{{\cal Q}}^2\lambda( {\cal P}\,\mathbf{
I} +\overline{{\cal P}}\,\overline{\mathbf{I}});\nonumber\\
\bthorn {\Lambda} &=&0, \quad \
\bedth {\Lambda}= 0, \quad
\bedthp {\Lambda} = 0, \quad
\bthornp {\Lambda} =0
\end{eqnarray}
We can solve in terms of Cartan spinor invariants for ${\cal Q}$ at zeroth order and for ${\cal P}$  at first order. Also, we can solve for  $( {\cal P}\,\mathbf{
I} +\overline{{\cal P}}\,\overline{\mathbf{I}})$ at first order; therefore ${\bf I}$ is not uniquely determined, and  it has  clearly the gauge freedom of a one parameter subgroup of null rotations,
\begin{eqnarray}\label{isotropy}
\mathbf{I}\ \to \ \mathbf{I} +i\varepsilon \bar {\cal P} {\cal Q}{\pmb o}
\end{eqnarray}
where $\varepsilon$ is an arbitrary real zero-weighted scalar.\footnote{this corrects a typo in equation (44) in \cite{edram2}.}

Since new information about the essential coordinates has arisen, we must go to the next order.

\noindent
\underbar{At second order}, a complete set of independent Cartan spinor invariants is
\begin{eqnarray}  \label{second}
 \bthorn \bthorn{\Phi} = 0, \quad
  \bedth \bthorn{\Phi} = 0, \quad
 \bthorn' \bthorn{\Phi} = 0, \quad \bedth' \bedth{\Phi} =  0, \nonumber\\
\bedth \bedth{\Phi} =  2\lambda^2{\cal Q}^2{\cal P}^2,\quad
\quad
\bthornp \bedth{\Phi} =- 3{{\cal Q}}^2\lambda^2 {\cal P} ( {\cal P}\,\mathbf{
I} +\overline{{\cal P}}\,\overline{\mathbf{I}}), \nonumber \\
\bthornp \bthornp {\Phi} =-{3}{\cal Q}^4\lambda^2 \Bigl(2  m \nu_5(r ) +1/\lambda^2\Bigr)  +12{{\cal Q}}^2\lambda^2\bigr({\cal P} \,  {\bf I  } +  \overline {\cal P} \, \overline{\bf I  }\bigl)^2
\end{eqnarray}
together  with complex conjugates. The GIF commutator equations
enable us to concentrate on this reduced list of independent
invariants.

We can replace the last expressions in (\ref{second}), using respectively the invariants from the left hand sides of (\ref{zeroth}) and last invariant in (\ref{first}), as
\begin{eqnarray}  \label{second+}
{ \X}=\Bigl(\bthornp\bthornp{\Phi}-\frac{4}{3}(\bthornp{\Phi})(\bthornp{\Phi})/{\Phi}\Bigr)\Big{/}{\Phi}^2=-3\lambda^2\Bigl( 2m\nu_5(r )+1/\lambda^2\Bigr) .
\end{eqnarray}
This highlights that part of the only second order invariant from which any new information must come. At this order the  zero-weighted scalar $\Bigl(  m \nu_5(  r) \Bigr)$ can be taken as the first  essential base  coordinate, in general, providing $\nu_5(r )\neq 0$. Furthermore,
 ${\bf I}$ is  still not uniquely determined, having still the gauge freedom of a one parameter subgroup of null rotations.

Since new information about the essential coordinates has arisen, we must go to the next order.

\noindent
\underbar{At third order},  we can see from the tables  in \cite{edram2} and the information already obtained at second order, that any new information could only  come from  operating on the last of the second order invariants $\Ph'\Ph' \Phi$, from which we obtain,

\begin{eqnarray}  \bedth {\X} &=&  -6\lambda^3{\cal P}m\nu_5(r ) \label{third1}\\
\bthornp {\X} &=&-6\lambda^2{\cal Q}m^{3/2}\Bigl( \nu_5'(  r)+\nu_5(  r)\nu_4(  r)\Bigr)
\label{third2}\end{eqnarray}
where prime, as before, indicates differentiation with respect to $r$.
The first of these equations (\ref{third1}) gives no new information, but from (\ref{third2}) the non-trivial scalar $\Bigl(m^{3/2}\bigl( \nu_5'(  r)+\nu_5(  r)\nu_4(  r)\bigr)\Bigr)$ is obtained; this scalar is  functionally independent of the first, in general. Hence, we can adopt$\Bigl(m^{3/2}\bigl( \nu_5'(  r)+\nu_5(  r)\nu_4(  r)\bigr)\Bigr) $ as a second essential base coordinate in general.

Furthermore, for all the third order Cartan spinor invariants, there is no new information about ${\bf I }$ and so the gauge freedom of ${\bf I }$
still remains unchanged.

Since new information about essential coordinates has arisen, we must go to the next order.

\noindent
\underbar{At fourth order}, we can see from the tables  in \cite{edram2} and the information already obtained at lower orders, that any potentially new information could only come from operating on (\ref{third2}) , which gives
\begin{eqnarray}
\bedth\bthornp {\X} &=& -9\lambda^3{\cal P}m^{3/2}\Bigl( \nu_5'(  r)+\nu_5(  r)\nu_4(  r)\Bigr) \label{fourth1}\\
\bthornp \bthornp {\X} &=&  -9\lambda^2{\cal Q}^2  m^{2}\Bigl( \nu_5'(  r)+\nu_5(  r)\nu_4(  r)\Bigr) \nu_4 (r)  \nonumber\\
&&-6\lambda^2{\cal Q}^2  m^{2}\Bigl( \nu_5''(  r)+\nu_5'( r)\nu_4(  r)+\nu_5( r)\nu_4'( r)\Bigr)
\label{fourth4}\end{eqnarray}
We find, {\it in general},  that there are no new functionally independent scalars generated; moreover, the gauge freedom of a one parameter subgroup of null rotations for ${\bf I  }$ (\ref{isotropy}) remains unchanged for  all fourth order spinor Cartan invariants.  Hence,  no new information about the essential coordinates has been obtained at this order; therefore  the algorithm terminates at fourth order, in gereral.

\medskip
\noindent
Moreover, in the discussions above for the general case, we have overlooked   two special cases which have to be considered separately:

\noindent
$\bullet$ the case where $m\nu_5(  r)$ and $ m^{3/2}\Bigl( \nu_5'(  r)+\nu_5(  r)\nu_4(  r)\Bigr)$ are functionally dependent, i.e., $ \nu_5'(  r)+\nu_5(  r)\nu_4(  r)=k \bigl(\pm\nu_5(  r)\bigr)^{3/2} $, where $k$ is a constant including zero;

\noindent
$\bullet$ the case
 when $\nu_5(  r) =0  $.

\medskip

\noindent
The results can be summarised as follows,

\smallskip
\noindent
{\bf (a)}\  When $\nu_5( r)\ne 0$  and  $ \nu_4(
r)\ne \Bigl(k\bigl(\pm\nu_5( r)\bigr)^{3/2}- \nu_5'(
r)\Bigr)/ \nu_5( r) $ for any constant $k$ (including zero), a first essential base coordinate is found at second order, a second essential base coordinate is found at third order, and no new essential base coordinate is obtainable at fourth order; the one degree of isotropy freedom at first order still remains up to fourth order. Since no new information on isotropy and essential base coordinates is given at fourth order, the procedure formally terminates at fourth order; therefore, this subclass has two essential base coordinates, which we can identify as $m$ and $r$ and there is one degree of isotropy freedom.

\smallskip
\noindent
{\bf (b)} \
When $\nu_5( r) \ne 0$ and $ \nu_4(
r)= \Bigl(k\bigl(\pm\nu_5( r)\bigr)^{3/2}- \nu_5'(
r)\Bigr)/ \nu_5( r) $ for any constant $k$ (including zero), a first essential base coordinate is found at second order, but no new essential base coordinate is given at third order. Since no new information on isotropy and essential base coordinates is given at third order, the procedure formally terminates at third order; therefore, this subclass has one essential base coordinate $ \bigl(m \nu_5(
r)\bigr)$and there is one degree of isotropy freedom.

\smallskip
\noindent
{\bf (c)}\
When $\nu_5( r)= 0$ there is a (partial) restriction on the null rotations at first order, but no new information on isotropy and essential coordinates is given at second order. Hence, the procedure formally terminates at second order; therefore, this subclass has no essential base coordinates, and there is one degree of isotropy freedom.

\subsection {Redundant and arbitrary  functions and simple metric forms}\label{redundancy}

{\bf Redundant functions}

\noindent
Two of the apparently arbitrary functions $\nu_3(r)$ and $ \nu_6( r) $,  which occur in the metric (\ref{metric1up}), have not occurred in the expressions for the Cartan spinor invariants, for any of the three cases; and since  the algorithm terminates at fourth order, there will be no subsequent information linking these functions with Cartan invariants.
This means that the apparent freedom of these  functions is not actual, and we can equate them to zero. This is confirmed when we consider the explicit coordinate transformation
$$ r\to r,\quad  n\to n +\beta(r)b +\alpha(r), \quad m \to m , \quad b \to b   +\gamma(r ) ,$$
which with appropriate choices of $\alpha(r)$, $\beta(r)$ $\gamma(r)$ enables us to make the explicit transformation $\nu_3( r)= 0$ and $ \nu_6( r) = 0 $, so that the spacetime (\ref{metric1up}), in the cases {\bf(a)}and {\bf (b)}, has the simpler form for $V$ given by

\begin{equation}\label{Usim3red}
V  ={3}  n\nu_4( r) -  {\nu_5} ( r)( b^2+
 m^2)  -\frac{ m}{\lambda^2}
\end{equation}
In addition in case {\bf (c)}, as well as $\nu_3( r)$ and $ \nu_6( r), $ the arbitrary function $\nu_4(r)$ does not occur in any of the Cartan invariants and so also can be equated to zero; this is confirmed by the additional  coordinate transformation
$$ r\to r,\quad  n\to n\mu(r), \quad m \to m \kappa(r ) ,\quad b \to b \mu(r),$$
where $\mu(r)= \exp(-\int^r \nu_4(s) ds), \ \kappa(r)=\mu(r)^{3/2}$, giving the very simple metric,
\begin{equation}
g^{ij}=  { m^2}\left(
\begin{array}{lccr}
0 &1& 0 & 0
\cr 1&
-m/\lambda^2 &  0&
 0
\cr 0 &   0&-{2}\lambda^2 & 0
 \cr 0 &
 0
 & 0 &-{2}\lambda^2\end{array}
\right)
\end{equation}

\begin{equation}
ds^2=\left( 2m dr^2+4\lambda^2dr dn-dm^2-db^2\right)\Big{/}2\lambda^2m^2 \label{metricN}
\end{equation}
This is the simplest conformally flat, pure radiation spacetime with a negative cosmological constant. It is the only case in the otherwise Type N family of homogeneous solutions of pure radiation spacetimes with a negative cosmological constant
\begin{equation}
ds^2=\left(\pm 2m^{2\tilde k} dr^2+4\lambda^2dr dn-dm^2-db^2\right)\Big{/}2\lambda^2m^2 \label{metric4down}
\end{equation}
given in equation (12.38) in (\cite{kramer}), and originally investigated in (\cite{sik}). (We have retained the slightly different coordinates used above, for easy comparison.) Other special cases of this family are the Kaigorodov spacetimes (\cite{kai}) when $\tilde k=3/2$, and the Defries spacetime (\cite{def}) when $\tilde k=-1$.

\noindent
{\bf Arbitrary functions}

\noindent
In cases {\bf (a)} and {\bf (b)}, we have the explicit occurrence of the functions $\nu_4(r )$ and $\nu_5(r )$ in the Cartan invariants; as to whether these are genuinely arbitrary functions can be determined from the information concerning their derivatives, which occur at the final step in each algorithm for cases {\bf (a)} and {\bf (b)} respectively.
In case {\bf (a)}, up to and including fourth order where the algorithm terminates, there are only three independent Cartan invariants which involve $\nu_4(r )$, $\nu_5 (r )$ and their first derivatives; since we identify two essential base coordinates from these, this leaves with only one invariant to give information about derivatives; hence only one function is truly arbitrary. In case {\bf (b)}, the algorithm terminates at third order and there is only one independent Cartan invariant since
$\bthornp {\X} =-6\lambda^2{\cal Q}(\pm m\nu_5(r))^{3/2}k$ and hence we can extract one essential base coordinate and one arbitrary constant $k$, but
no arbitrary function.

\subsection {Killing and homothetic vectors}

Siklos \cite{sik} has pointed out that the calculations for the symmetries  in his (more general--- Type N as well as Type O, radiation) spacetime were quite difficult. On the other hand, our symmetry calculations in \cite{edram4} were very simple, because  the version of the spacetime from \cite{edvic} which we used was built from an intrinsic GHP tetrad.  Unfortunately, the spacetime which we are now considering does not have an  intrinsic GHP tetrad, and therefore the very simple results derived in \cite{edlud3} which we exploited  in \cite{edram4} are not valid in the present case. However, the modification to the procedure in  \cite{edlud3}, which we outlined in the Introduction, for spacetimes with null rotation isotropy freedom,  can be used. 

From the classification procedure just completed we have deduced that in the spacetime under discussion there is one degree of null rotation isotropy, (\ref{isotropy}), and so, taking into account the comments in the Introduction, we introduce the new family of tetrads $\widehat {\bf l}$, $\widehat {\bf n}$, $\widehat {\bf m}$, $\widehat {\bf \bar{m}}$ as given by eq. (\ref{newbasis}) with $z=i\cal{P}\cal{Q}\varepsilon$, where the real zero-weighted parameter $\varepsilon$ gives the permitted one degree of isotropy freedom.
From the new basis (\ref{newbasis}), the definitions of the GHP operators and the transformation rules for the spin coefficients (see \cite{ghp}),
the GHP operators transform as
\begin{eqnarray}\nonumber
\widehat {\thorn} &=&{\thorn}; \quad \widehat {\edth} =\edth +i{\cal P}{\cal Q} \varepsilon \thorn; \quad \\\label{GHPtransformation}
\widehat{\thorn'}&=&\thorn'-i\bar{\cal{P}}{\cal Q}\varepsilon\edth +i{\cal{P}}{\cal Q}\varepsilon \edth'+ {\cal Q}^2\varepsilon^2\thorn+i(p-q){\cal Q}\lambda\varepsilon
\label{hatoperators}
\end{eqnarray}
when acting on objects of weight $\{p,q\}$.

In \cite{edram2}, even though a second intrinsic spinor was not generated by the analysis, it was necessary to introduce a second spinor in order to transfer to the GHP formalism
and hence to complete the integration procedure; this resulted in the GHP tables  (71), (72), (91) - (98)  in \cite{edram2}.
These   tables  can be easily translated into tables for the new
operators (\ref{GHPtransformation}.
We can then  use the GHP Lie derivative operator (\ref{GHPLhat}) and Lie-GHP commutators (\ref{commhat}) with respect to these new operators.

\smallskip

From their definitions in \cite{edram2}, the weighted scalars ${\cal P}$, ${\cal Q}$ are both clearly intrinsic GHP scalars; furthermore, it is easy to see that in this particular class of spacetimes, these particular GHP scalars are invariant under null rotations. Hence $\cal{P}$ and $\cal{Q}$ are also GHP scalars with respect to any null tetrad in the family which differs by a null rotation from the original tetrad. Since we choose to use a new tetrad, which by definition is GHP Lie derived in the presence of a Killing vector, and is within this family, then $\cal{P}$ and $\cal{Q}$ are also GHP Lie derived by Lemma 2, \cite{edlud3}, i.e.,

\begin{equation}
\widehat{\hbox{\L}}_\xi{\cal P}=\ 0\ =\widehat{\hbox{\L}}_\xi{\cal Q}.
\label{LhatPQ}
\end{equation}

Since the GHP Lie derivative for zero weighted scalars reduces to the usual Lie derivative we have
that the components of the Killing vectors in the (zero-weighted) coordinates $r$, $n$, $m$,$b$ are
$\xi^{r}\equiv{\cal L}_\xi r=\widehat{\hbox{\L}}_\xi r$ etc.

From their definitions in \cite{edram2}, none of the four (zero-weighted) coordinates $r$, $n$, $m$, $b$ is a GHP scalar, and so, in general, none will be GHP Lie derived by any Killing vectors present; rather they are complementary scalars. Therefore, in order to obtain explicit expressions for the Killing vectors we have to solve the GHP-Lie commutators (\ref{commhat}) applied to each of these coordinates.

Application of (\ref{commhat}), with the transformed operators given by (\ref{GHPtransformation}), gives

\begin{eqnarray}\nonumber
 \thorn \xi^{m} &=& 0, \quad \edth \xi^{m}  =   \lambda{\cal{P}}\xi^{m}, \quad
\thorn' \xi^{m} = {\cal{Q}}m^{1/2}\left(\frac{3}{2}\nu_4\xi^{m} +
m\nu_4'\xi^{r}\right)\\\nonumber
 \thorn \xi^{r} &=& 0, \quad \edth \xi^{r}  =   0,
\quad
\thorn' \xi^{r} =\frac{1}{2}{\cal{Q}}m^{-1/2}\xi^{m}\\\nonumber
\thorn \xi^{b} &=& 0, \quad \edth \xi^{b}  =   i\lambda{\cal{P}}\xi^{m},
\nonumber\\\nonumber
\thorn' \xi^{b} &=&{\cal{Q}}m^{1/2}\left( \frac{1}{2}m^{-1}\nu_4b\xi^{m} + \nu_4\xi^{b} + \nu_4'b\xi^{r}
+2\lambda m^{1/2}\widehat{\hbox{\L}}_\xi\varepsilon\right)\\\nonumber
 \thorn \xi^{n} &=&{\frac{3}{2{\cal{Q}}}}m^{1/2}\xi^{m}, \quad \edth \xi^{n}  =   i{\cal{P}}m^{3/2}\widehat{\hbox{\L}}_\xi\varepsilon, \quad
\\
\thorn' \xi^{n} &=& \frac{1}{2}m^{-1/2}{\cal{Q}}\left(m\widehat{\hbox{\L}}_\xi V + \frac{1}{2}V\xi^{ m}\right)
\label{ComLn}
\end{eqnarray}

\smallskip
\noindent
Using the explicit form of the GHP operators, remembering that the $\xi^\mu$ are zero-weighted
so that ${\thorn}=l^a\nabla_a$ etc., these translate into

\begin{eqnarray}\nonumber
 \xi ^{m}_{,n} &=& 0, \quad \xi^{m }_{,m}  = m^{-1}  \xi^{m}, \quad
  \xi ^{m }_{,b} = 0,
\quad
  \xi ^{m }_{,r} = \frac{1}{2}\nu_4\xi ^{m} + m\nu_4'\xi ^{r}\\
\nonumber
 \xi^{ r }_{,n} &=& 0, \quad \xi^{r }_{,m} =   0 ,
\quad  \xi^{r}_{,b} =   0, \quad 
\xi ^{r }_{,r} =\frac{1}{2}m^{-1}\xi ^{m}\\
\xi ^{b }_{,n}&=& 0, \quad \xi^{b  }_{,m}=   0, \quad \xi^{b}_{,b}=m^{-1}\xi^{ m}, \quad \nonumber\\\nonumber
\xi ^{b }_{,r} &=& -\frac{1}{2}m^{-1}\nu_4b\xi^{ m} + \nu_4\xi^{b} + \nu_4'b\xi ^{r} +2\lambda m^{1/2}\widehat{\hbox{\L}}_\xi\varepsilon\\\nonumber
\xi ^{n }_{,n}&=&\frac{3}{2}m^{-1}\xi ^{m}, \quad \xi^{n }_{,m} =  0, \quad \xi ^{n }_{,b}=\frac{m^{1/2}}{\lambda}\widehat{\hbox{\L}}_\xi \varepsilon,\nonumber\\
\xi ^{n }_{,r}&=& -\frac{1}{2}m^{-1}V\xi ^{m} + \frac{1}{2}\widehat{\hbox{\L}}_\xi V-\nu_4 b\frac{m^{1/2}}{\lambda}\widehat{\hbox{\L}}_\xi \varepsilon ,
\label{ComLn1}
\end{eqnarray}
where $\widehat{\hbox{\L}}_\xi V=\left(3n\nu_4'-\nu_5'(b^2+m^2)-\nu_6'b+\nu_3'\right)\xi^r+3\nu_4\xi^n
-\left(2\nu_5 m+\frac{1}{\lambda^2}\right)\xi^m-\left(2\nu_5b+\nu_6\right)\xi^b$.

The solution to this system, i.e. the Killing vectors, is then in terms of the functions $f\equiv f(r)$, $g\equiv g(r)$ and $h\equiv h(r)$ given by

\begin{eqnarray}\label{killingvectors}\nonumber
\xi^{m}&=&{m}f^\prime\; , \quad \xi^{r}=\frac{1}{2}f\; ,
\quad \xi^{b}={b}f^\prime+g\; , \\
\quad \xi^{n}&=&\frac{3}{2}{n}f^\prime+\frac{b}{2\lambda^2}
\left(g^{\prime}-\nu_4 g\right)+h
\end{eqnarray}
where $g$ satisfies
\begin{equation}\label{g}
g^{\prime\prime}-\frac{3}{2}\nu_4 g^{\prime}+\left(2\lambda^2\nu_5-\nu_4^\prime+\frac{1}{2}\nu_4^2\right)g=
-\frac{1}{2}\lambda^2\nu_6^{\prime}f \, ,
\end{equation}
which gives two constants of integration,
and $h$ is given by
\begin{eqnarray}\nonumber
&&h(r)=\exp{\left(\frac{3}{2}\int^r\nu_4(s)ds\right)}\Bigl[C_1+\Bigr.\\
&&\Bigl. \int^r \exp\left(-\frac{3}{2}\int^s \nu_4(t) dt\right)
 \left(-\frac{1}{2}\nu_3(s)f^\prime+\frac{1}{4}\nu_3^{\prime}(s)f-\frac{1}{2}\nu_6(s)g\right) ds\Bigr]
\end{eqnarray}
with $C_1$ a constant of integration. Finally $f$ should satisfy the two equations
\begin{equation}\label{constraint}
f^{\prime\prime}-\frac{1}{2}(\nu_4 f)^{\prime}=0 \quad \hbox{and} \quad
\nu_5 f^{\prime}+\frac{1}{2}\nu_5^{\prime} f=0 \, .
\end{equation}
In the following we will use the coordinate freedom, as discussed in section \ref{redundancy}, to put the functions $\nu_3$ and $\nu_6$ to zero.

\subsubsection{Killing vectors in the generic case (a)}

In this case, with arbitrary $\nu_4$ and $\nu_5$, (\ref{constraint}) implies that $f\equiv 0$.
Hence
\begin{equation}
\xi^{m}=\xi^{r}=0\; , \quad \xi^{b}=g\; , \quad \xi^{n}=\frac{b}{2\lambda^2}
\left(g^{\prime}-\nu_4 g\right)+h
\end{equation}
and there are three constants of integration and three independent Killing vectors. It is hard to find the
general solution to (\ref{g}), but the function $A$ in (\ref{Hu}) can be made equal to zero through a
coordinate transformation, see \cite{sik}. Hence, we can by (\ref{ABCD}) choose
\begin{equation}\label{A}
A=2\lambda^2\nu_5-\nu_4^\prime+\frac{1}{2}\nu_4^2=0\, .
\end{equation}
Equation (\ref{g}) is then easily integrated to give
\begin{equation}
g(r)=C_2\int^r\exp{\left(\frac{3}{2}\int^s\nu_4(t)dt\right)}ds+C_3 \, .
\end{equation}
The three independent Killing vector are then
\begin{eqnarray}\nonumber
\xi_{(1)}&=&\exp{\left(\frac{3}{2}\int^r\nu_4(s) ds\right)}\frac{\partial}{\partial n} \\\nonumber
\xi_{(2)}&=&\frac{b}{2\lambda^2}\exp{\left(\frac{3}{2}\int^r\nu_4(s) ds\right)}\frac{\partial}{\partial n}+\\\nonumber
&&\int^r\exp{\left(\frac{3}{2}\int^s\nu_4(t) dt\right)}ds\left(\frac{\partial}{\partial b}-\frac{b\nu_4}{2\lambda^2}
\frac{\partial}{\partial n}\right) \\
\xi_{ (3)}&=&-\frac{b\nu_4}{2\lambda^2}\frac{\partial}{\partial n}+\frac{\partial}{\partial b}\, .
\end{eqnarray}
Using the coordinate transformation (\ref{coordtransform}) these are (up to allover constant factors)
transformed to
\begin{eqnarray}\label{killingcasaa}
\xi_{(1)}=\frac{\partial}{\partial v}\, , \quad 
\xi_{(2)}=y\frac{\partial}{\partial v} -u\frac{\partial}{\partial y}\, , \quad
\xi_{(3)}=\frac{\partial}{\partial y}\,
\end{eqnarray}
which are the Killing vectors $I$, $V$ and $IV$ in Siklos' paper \cite{sik}. This case corresponds to the fifth row in
his table 1 with his $A(x,u)=f(u)x$.

\subsubsection{Killing vectors in the case b}

This case is given by $\nu_5 \ne 0$ and
\begin{equation}\label{k}
 \nu_4=-\frac{\nu_5'}{ \nu_5}+k\left(\pm \nu_5\right)^{1/2}
\end{equation}
for any constant $k$ (including zero).
The function $f$ is then given by
\begin{equation}
f=C_4\left(\pm\nu_5\right)^{-1/2}
\end{equation}
where $C_4$ is a constant of integration. Hence, from (\ref{killingvectors}), the fourth Killing vector is
\begin{equation}
\xi_{(4)}=\left(\pm\nu_5\right)^{-1/2}\frac{\partial}{\partial r}\mp\frac{\nu_5^\prime}{\left(\pm\nu_5\right)^{3/2}}
\left[m\frac{\partial}{\partial m}+\frac{3}{2}n\frac{\partial}{\partial n}+b\frac{\partial}{\partial b}\right]\, .
\end{equation}
As can be seen from an invariant classification, a general enough solution to (\ref{k}) is given by
\begin{equation}\label{D1}
\nu_5=\pm\frac{D_1^2}{r^2}\, , \quad \nu_4=\frac{2+kD_1}{r}
\end{equation}
where $D_1\geq 0$ is a constant. From (\ref{coordtransform}) one then obtains
\begin{equation}\label{killing4}
\xi_{(4)}=\left(\frac{3}{2}kD_1+4\right)\frac{\partial}{\partial u}
-\frac{3}{2}kD_1\frac{\partial}{\partial v}+2x\frac{\partial}{\partial x}+2y\frac{\partial}{\partial y}\, ,
\end{equation}
which agrees with $VI_\alpha$ in \cite{sik} with $\alpha=-\frac{3}{2}kD_1-2$.

The gauge (\ref{A}) can be made consistent with the choice (\ref{D1}) for $k<2\lambda$, so that the
three other Killing vectors are still given by (\ref{killingcasaa}). This case corresponds to row 11 in
table 1 in \cite{sik} with $u^{-2\beta-2}A(xu^\beta)=xu^{-\beta-2}$ and $\alpha=2(1+1/\beta)$.

If $k\ge 2\lambda$ (\ref{D1}) cannot be made consistent with (\ref{A}) with a real $D_1$, but equation (\ref{g})
is now easily integrated with an ansatz $r^\alpha$. The Killing vectors $\xi_{(2)}$ and
$\xi_{(3)}$ now get slightly more complicated than the corresponding ones in (\ref{killingcasaa}), but it
is straightforward to show that the four Killing vectors satisfy the same Lie algebra as (\ref{killingcasaa})
and (\ref{killing4}) do.

\subsubsection{Killing vectors in the spacetime homogeneous case c}

When $\nu_5\equiv 0$ the spacetime becomes homogeneous and also, as in the generic case,
has one isotropy.
As was discussed in section \ref{redundancy} $\nu_3$, $\nu_6$  and $\nu_4$ never appear in
the classification and they may all  be put to zero. With $\nu_3=\nu_6=\nu_4=0$
equations   are easily solved to give
\begin{equation}
f=C_4 r+C_5 \; , \quad g=C_2 r+C_3 \; , \quad h=C_1 \;
\end{equation}
so that
\begin{eqnarray}\nonumber
\xi^{m}&=&C_4{m}\; , \quad \xi^{r}=\frac{1}{2}\left(C_4{r}+C_5\right)\; ,
\quad \xi^{b}=C_4{b}+C_2{r}+C_3\; , \\
\quad \xi^{n}&=&\frac{3}{2}C_4{n}+\frac{C_2b}{2\lambda^2}
+C_1 \; .
\end{eqnarray}
The five independent Killing vectors are hence 
\begin{eqnarray}\nonumber
\xi_{(1)}&=&\frac{\partial}{\partial v}\, , \quad
\xi_{(2)}=\frac{b}{2\lambda^2}\frac{\partial}{\partial v}+r \frac{\partial}{\partial b}\, ,
\quad
\xi_{(3)}=\frac{\partial}{\partial b}\, , \\\label{Killing5}
\xi_{(4)}&=&\frac{r}{2}\frac{\partial}{\partial r}+\frac{3n}{2}\frac{\partial}{\partial n}+
m \frac{\partial}{\partial m}+b \frac{\partial}{\partial b}\, , \quad
\xi_{(5)}=\frac{\partial}{\partial r}
\end{eqnarray}
Using the coordinate transformation in (\ref{coordtransform}) the vectors (\ref{Killing5}) then transform
 (up to overall constant factors) into
\begin{eqnarray}\nonumber
\xi_{(1)}&=&\frac{\partial}{\partial v}\, , \quad \xi_{(2)}=y\frac{\partial}{\partial v}-u\frac{\partial}{\partial y}\, ,
\quad \xi_{3}=\frac{\partial}{\partial y}\, , \\\label{Killing2}
\xi_{(4)}&=&u\frac{\partial}{\partial u}+3v\frac{\partial}{\partial v}+2x\frac{\partial}{\partial x}+2y\frac{\partial}{\partial y}\; ,\quad
\xi_{(5)}=\frac{\partial}{\partial u}\; .
\end{eqnarray}
These agree
with the Killing vectors  $I$,  $V$,  $IV$, $VI_1$ and $III$, i.e., row 12 with $\alpha=1$  in table 1 in \cite{sik}.

\subsubsection{Homothetic Killing vectors}

A method for determining homothetic vectors for spacetimes constructed from GHP formalism is given in \cite{edlud10}. As in the procedure given in \cite{edlud3} for determining Killing vectors, this method is valid when two null directions are defined geometrically, which is not the case here, since the GHP tetrad has one degree of isotropy freedom. We can, however, generalise the results given in \cite{edlud10} to the study of homothetic vectors when only one null direction is singled out in a similar way we have generalized the method for Killing vectors in section 1.3.
All GHP scalars are 
properly conformally weighted when the conformal factor $\varphi$
is constant, i.e. for Killing and homothetic vectors. Hence, for ${\cal P}$ and ${\cal Q}$ we can write:
\begin{eqnarray}
\widehat{\hbox{\L}}_\xi{\cal P}&=&0,\label{LhatP}\\
\widehat{\hbox{\L}}_\xi{\cal Q}&=&-\frac{\varphi}{2}Q\label{LhatQ}
\end{eqnarray}
where $\xi$ is a homothetic vector and ${\cal P}$ and ${\cal Q}$ have conformal weight equal to 0 and -1 respectively.

By theorem 1 in \cite{edlud10} both ${\cal P}$ and ${\cal Q}$ must satisfy the commutator:
\begin{equation}
 \Bigl[ \widehat{\hbox{\L}}_\xi,
\widehat{\thorn}\Bigr]\eta=-\frac{\varphi}{2}\widehat{\thorn}\eta, \label{commhom}
\end{equation}
and its companion equations when $\varphi$ is constant.
A simple calculation shows that for
\begin{equation}
 \Bigl[ \widehat{\hbox{\L}}_\xi,
\widehat{\edth}\Bigr]{\cal P}=-\frac{\varphi}{2}\widehat{\edth} {\cal P}, \label{commhom}
\end{equation}
to be satisfied we must have $\varphi=0$, so that we can quickly conclude that no homothetic vectors are present.

\section {Summary and Discussion}

GIF has been used to investigate CFPR spacetimes and to carry out an invariant classification and investigation of the equivalence problem for a subclass of CFPR spacetimes \cite{edram2}, \cite{edram3} (the case with zero cosmological constant was analysed in \cite{edvic}). Because of the inherent simplicity of the invariant classification in GIF the analysis in this paper illustrates explicitly some of the more subtle aspects of the invariant classification scheme and investigation of the equivalence problem.

According to the Cartan formulation of the equivalence problem, successive orders of the Riemann tensor and its derivatives are considered until at some order the information obtained at the highest order should reveal no new information about the isometry group nor the number of essential coordinates; on the other hand, this step  may reveal information about the functional independence of arbitrary functions, through other Cartan invariants. In the case studied here we were able to prove that at most only one of the apparently arbitrary functions is truly arbitrary and in some cases none of the seemingly arbitrary functions are in fact arbitrary.

As in \cite{edvic} and \cite{edram2}, having constructed the spacetime via GIF, we find it is easy to deduce the Karlhede classification; also, as in \cite{edvic} and \cite{edram2}, the Karlhede algorithm required the fourth order in the derivatives of the Riemann tensor. The fact that, for these two classes of spaces, we can carry out the Karlhede classification \textit{by hand} as a simple calculation rather than by the more complicated spinor calculations associated with the computer programmes for the Karlhede algorithm emphasises the power of the method. Moreover, we were able to
see directly how different aspects of the Karlhede algorithm, especially regarding null isotropy and multiple Killing vectors, manifested themselves in the GIF.

We noted in \cite{edram2} that, in \cite{edvic} we could have simplified the Karlhede classification calculation, by changing
from GIF operators to the simpler GHP scalar operators from the beginning; this is permissable in \cite{edvic} because the second dyad spinor  ${\setbox0=\hbox{$\iota$} 
\kern-.025em\copy0\kern-\wd0 \kern.05em\copy0\kern-\wd0 \kern-0.025em\raise
.0433em\box0 }(\equiv \mathbf{I})$, which enables us to translate from GHP formalism to GIF, is intrinsic and invariant in the GIF. On the contrary,
for the spacetimes in \cite{edram2}, we have seen that we do not get an intrinsic second spinor from the GIF formalism; rather the spinor $\mathbf{I} $ which we used has one degree of freedom fixed in a non-intrinsic manner, and so such a change from GIF to GHP in the Karlhede classification was not permitted. For the class of spacetimes investigated in this paper, we also obtained a second unique intrinsic spinor, and so a change  to GHP from the beginning of the Karlhede classification would have been permissable, although we did not make it. However, in such a situation, we would need to keep in mind at which level the second spinor was fixed, in order to keep track of each step and condition of the Karlhede algorithm.

Unfortunately, the spacetime which we are now considering does not have an intrinsic GHP tetrad, and therefore the very simple results derived in \cite{edlud3} which we exploited in \cite{edram3} are not valid in the present case. By generalising the procedure in \cite{edlud3} to the case of the spacetimes described here, which does not have an intrinsic GHP tetrad, we were able to determine the Killing vectors and compare them to the results obtained by Siklos in \cite{sik}. Our study shows that this approach is simpler and more efficient than the usual procedure as used by Siklos in \cite{sik}.

Furthermore, in a similar fashion to the way we treat Killing vector analysis for these spacetimes, we generalise the method described in \cite{edlud10}, for determining homothetic vectors in GHP formalism. In doing so, it is very easy to see that there are no existing homothetic vectors.

\section*{Appendix}

\appendix\label{Karlhede}
\section{Cartan-Karlhede classification by CLASSI}

For comparison the metrics (\ref{metric1}) and (\ref{metric1.1down}) are here classified according to the Cartan-Karlhede procedure
using the program package CLASSI \cite{classi} and the condition for equivalence between the two metrics is given.

\subsection{Classification of the metric in GIF coordinates}\label{classint}

The null tetrad (\ref{frame}) found in the GIF procedure, being as intrinsic as possible, should be  suitable
 for classifying the metric (\ref{metric1}). If we also keep the one-dimensional isotropy freedom given by $\varepsilon$ in (\ref{hatoperators}),
the corresponding 1-form basis is
\begin{eqnarray}\nonumber
\omega^0&=&\frac{{\cal{Q}}}{m^{3/2}}\left(-\frac{1}{2}Vdr+dn\right)+
\frac{{\cal{Q}}\varepsilon^2}{m^{1/2}}dr
+\frac{{\cal{Q}}\varepsilon}{\lambda m}\left(b\nu_4dr-db\right) \, , \quad
\omega^1=\frac{1}{{\cal{Q}}m^{1/2}}dr
 \\\nonumber
\omega^2&=&\frac{\overline{{\cal{P}}}}{2\lambda m}\left(\nu_4(ib-m)dr+dm-idb\right)+
i\frac{{\overline{\cal{P}}}\varepsilon}{m^{1/2}}dr, \\
\omega^3&=&\frac{{\cal{P}}}{2\lambda m}\left(-\nu_4(ib+m)dr+dm+idb\right)-
i\frac{{\cal{P}}\varepsilon}{m^{1/2}}dr
\end{eqnarray}
where $V$ is given by (\ref{Usim3}).
The freedom in the GHP-parameters ${\cal{P}}$ and ${\cal{Q}}$ is kept to get the Riemann tensor and
its derivatives in a form such that
the functional dependence is minimised at each step. With the choices
${\cal{Q}}= 1$ and ${\cal{P}}=1$ 
the derivatives up to second order are given by
\footnote{Usually a standard frame such that $\Phi_{22^{\prime}}=+1$ is used.}
\begin{eqnarray}\nonumber
R&=&24\lambda^2, \;\;\;
\Phi_{22^{\prime}}=-1, \;\;\;
D\Phi_{23^{\prime}}=-\lambda, \;\;\; \\
D^2\Phi_{24^{\prime}}&=&\frac{4}{3}D^2\Phi_{33^{\prime}}=-2\lambda^2, \;\;\;
D^2\Phi_{44^{\prime}}=3+6\lambda^2\tilde s
\end{eqnarray}
where $\tilde s\equiv m\nu_5(r)$ is the first functionally independent quantity (if  $\nu_5\neq 0$). This form is invariant
under 1-dimensional null rotations.
The next functionally independent quantity, $\tilde w\equiv \nu_4/\nu_5^{1/2}+\nu_5^{\prime}/\nu_5^{3/2}$ is found in the third derivative
\footnote{We here only give the classification for $\nu_5>0$. The other case is treated similarly.}
\begin{eqnarray}\nonumber
D^3\Phi_{25^{\prime}}&=&-6\lambda^3, \quad
D^3\Phi_{34^{\prime}}=-\frac{18}{5}\lambda^3, \quad
D^3\Phi_{45^{\prime}}=18\lambda^3 \tilde s+\frac{42}{5}\lambda, \\
D^3\Phi_{55^{\prime}}&=&6\lambda^2 \tilde s^{3/2}\tilde w
\end{eqnarray}
which still is invariant under 1-dimensional null rotations.

No new coordinates are found in the 4:th derivative and the 1-dimensional isotropy remains.
Hence the 4:th derivative is sufficient for a complete classification
\footnote{In general the symmetrised derivatives are not sufficient for a complete classification.
The additional quantities needed are given in \cite{maccullum}. For the present case they do not
give any additional information.}.
The only 4:th derivative component with new information is
\begin{equation}
D^4\Phi_{66^{\prime}}=-180\lambda^4 \tilde s^2+9\lambda^2 \tilde s^2 \tilde w^2-
174\lambda^2 \tilde s-42+6\lambda^2 \tilde s^2 \tilde f(\tilde w)
\end{equation}
where $\tilde f(\tilde w)\equiv\frac{d \tilde w}{d r}/\nu_5^{1/2}$ is an arbitrary function of $\tilde w$ (or $r$).
Hence the metric has two essential coordinates, $\tilde s$ and $\tilde w$, a 1-dimensional isotropy group
and one arbitrary function, $\tilde f(\tilde w)$,
in the generic case.

If $\nu_5=0$ the metric is spacetime homogeneous with a 5-dimensional isometry group and the
classification ends at second order. When $\nu_4\nu_5+\nu_5'=k\nu_5^{3/2}$ for some constant
$k$, there is one essential coordinate and a 4-dimensional isometry group and the classification
ends at third order.
	
Notice that coordinates $\tilde s,\tilde w,\tilde x,\tilde y$ can be found where the metric
$ds^2=2\omega^0\omega^1-2\omega^2\omega^3$ is given by the 1-forms:
\begin{eqnarray}\nonumber
\omega^0&=&\big[\frac{\tilde s}{\lambda^2}+\tilde x^2+\tilde s^2-
3\tilde y \tilde w\big]\frac{d\tilde w}{2\tilde s^{3/2}\tilde f(\tilde w)}+
\frac{d\tilde y}{s^{3/2}},
\quad \omega^1=\frac{d\tilde w}{\tilde s^{1/2}\tilde f(\tilde w)} \\\nonumber
\omega^2&=&\frac{1}{2\lambda \tilde s}\big[d\tilde s-id\tilde x+
(i\tilde x-\tilde s)\tilde w\frac{d\tilde w}{\tilde f(\tilde w)}\big], \quad \omega^3=\overline{\omega^2}\, ,
\end{eqnarray}
clearly showing that there is at most one arbitrary function.

\subsection {Classification of the metric in Siklos coordinates}\label{classsik}

In the frame
\begin{eqnarray}\label{frameSiklos}\nonumber
\omega^0&=&-\frac{1}{4\lambda^2 x^2 z}(Hdu+2dv)+\eta^2zdu-\frac{\eta}{\lambda x}dx\; , \;\;
\omega^1=zdu\; , \\
\omega^2&=&\frac{1}{2\lambda x}(dx+idy)-\eta zdu\; , \;
\omega^3=\overline{\omega^2},
\end{eqnarray}
where $z$ and $\eta$ are used to get ${\cal R}^{q+1}$ in standard form, the Weyl spinor becomes
\begin{equation}
\Psi_4=\frac{1}{4z^2}(H_{,xx}-H_{,yy}-2iH_{,xy}) \; ,
\end{equation}
i.e., the metric is in general of Petrov type N. The solution to $\Psi_i=0$ is hence given
by (\ref{Hu}), verifying the result in \cite{pod2}.

For $B=0$ the metric corresponds to the anti de-Sitter spacetime and hence we assume $B\neq 0$
in the following.

A classification
with $z=\sqrt{-B/2x}$ and
$\eta=\frac{(2x)^{1/2}}{6\lambda(- B)^{3/2}}B_{,u}$ now gives
\footnote{We here choose $B$ to be negative (and $x$ positive) in accordance with (\ref{ABCD}).
A similar analysis can be made with the opposite sign of $B$.}
\begin{eqnarray}\nonumber
R&=&24\lambda^2, \;\;\;
\Phi_{22^{\prime}}=-1, \;\;\;
D\Phi_{23^{\prime}}=-\lambda, \;\;\; \\
D^2\Phi_{24^{\prime}}&=&\frac{4}{3}D^2\Phi_{33^{\prime}}=-2\lambda^2, \;\;\;
D^2\Phi_{44^{\prime}}=3+6\lambda^2s
\end{eqnarray}
where
\begin{displaymath}
s\equiv\psi(u)x\equiv\frac{1}{\lambda^2}\left(\frac{A}{B}+\frac{B_{,uu}}{3B^2}-
\frac{4}{9}\frac{B_{,u}^2}{B^3}\right)x
\end{displaymath}
and
\begin{eqnarray}\nonumber
D^3\Phi_{25^{\prime}}&=&-6\lambda^3, \quad
D^3\Phi_{34^{\prime}}=-\frac{18}{5}\lambda^3, \quad
D^3\Phi_{45^{\prime}}=18\lambda^3 s+\frac{42}{5}\lambda, \\
D^3\Phi_{55^{\prime}}&=&6\lambda^2 s^{3/2}w
\end{eqnarray}
where
\begin{displaymath}
w\equiv w(u)\equiv\left(-\frac{2}{B\psi}\right)^{1/2}\left(\frac{\psi_{,u}}{\psi}-\frac{B_{,u}}{3B}\right)
\end{displaymath}
The only 4:th derivative component with new information is
\begin{equation}
D^4\Phi_{66^{\prime}}=-180\lambda^4 s^2+9\lambda^2 s^2 w^2-174\lambda^2 s-42
+6\lambda^2 s^2 f(w)   \, ,
\end{equation}
where
\begin{displaymath}
f(w)\equiv \sqrt{-\frac{2}{\psi B}} w_{,u}
\end{displaymath}
is an arbitrary function of $w$ (or $u$).

If $\psi = 0$ the metric is spacetime homogeneous with a 5-dimensional isometry group and the classification
ends at second order and when $w$ is constant there is only one essential coordinate, four killing vectors and
the classification ends at third order. When $B=0$ the anti de-Sitter spacetime is obtained and the classifications
ends already at zeroth order. This case is not included in the metric (\ref{metric1}).

A comparison with the classification in appendix \ref{classint}
now shows that the two metrics (\ref{metric1.1down})
and (\ref{metric1}) are locally equivalent iff
\begin{eqnarray}\nonumber
\tilde s&\equiv&m\nu_5= \frac{1}{\lambda^2}\left(\frac{A}{B}+\frac{B_{,uu}}{3B^2}-\frac{4}{9}\frac{B_{,u}^2}{B^3}\right) x
\equiv\psi x\equiv  s
\; , \\\nonumber
\tilde w&\equiv&\frac{\nu_4}{\nu_5^{1/2}}+\frac{\nu^\prime_5}{\nu_5^{3/2}}=
\left(-\frac{2}{B\psi}\right)^{1/2}\left(\frac{\psi_{,u}}{\psi}-\frac{B_{,u}}{3B}\right)\equiv w\; , \\
\tilde f(\tilde w)&\equiv&\frac{d\tilde w}{dr}\frac{1}{\nu_5^{1/2}}=\left(-\frac{2}{B\psi}\right)^{1/2}
 w_{,u}
\equiv f(w)
\end{eqnarray}

\medskip

{\bf Acknowledgements}
This research was partially supported by the Research Center of Mathematics
of the University of Minho through the FEDER Funds Programa Operacional
Factores de Competitividade COMPETE, and by the Portuguese Funds
through FCT - Funda\c{c}\~{a}o para a Ci\^{e}ncia e Tecnologia within the Project Est-
C/MAT/UI0013/2011.
MB wishes to thank the Center of Mathematics of the University of Minho,
for supporting a visit to this university, and the Department of Mathematics and Applications for their
hospitality.



\end{document}